\documentclass[prd,preprint,superscriptaddress,amsmath,amssymb]{revtex4}

\usepackage{graphicx,color,slashed}

\begin{document}

\title{\bf  $\epsilon'/\epsilon$ from charged-Higgs-induced  gluonic dipole operators}

\author{Chuan-Hung Chen}
\email{physchen@mail.ncku.edu.tw}
\affiliation{Department of Physics, National Cheng-Kung University, Tainan 70101, Taiwan}

\author{Takaaki Nomura}
\email{nomura@kias.re.kr}
\affiliation{School of Physics, KIAS, Seoul 02455, Korea}

\date{\today}

\begin{abstract}
We study the effect of charged-Higgs-induced chromomagnetic operator, $Q_{8G}(-) \equiv \bar s \sigma^{\mu \nu} T^a \gamma_5 d G^a_{\mu \nu}$, on the Kaon direct CP violation $Re(\epsilon'/\epsilon)$.
Using the matrix element $\langle \pi \pi | O_{8G}(-) | K^0 \rangle$ recently obtained by a large $N_c$ dual QCD approach, we find that if the Kobayashi-Maskawa phase is the origin of CP violation, the charged-Higgs-induced gluon penguin dipole operator in the type-III two-Higgs-doublet model can explain the measured $Re(\epsilon'/\epsilon)$ when the constraints from the relevant low energy flavor physics, such as $\Delta B(K)=2$, $B\to X_s \gamma$, $B_s\to \mu^+ \mu^-$, and Kaon indirect CP violation parameter $\epsilon$, are included. 

\end{abstract}

\maketitle

Significant recent progress in meeting the
 long-standing challenge for the prediction of  the Kaon direct CP violation $Re(\epsilon'/\epsilon)$ in the standard model (SM) has been made  based on two results: (i) RBC-UKQCD collaboration obtains a surprising lattice QCD result on $Re(\epsilon'/\epsilon)$ with~\cite{Blum:2015ywa,Bai:2015nea}:
 \begin{equation}
 Re(\epsilon'/\epsilon) = 1.38(5.15)(4.59) \times 10^{-4}\,,
 \end{equation}
where the numbers in brackets denote the errors. (ii) Using a large $N_c$ dual QCD~\cite{Buras:1985yx,Bardeen:1986vp,Bardeen:1986uz,Bardeen:1986vz,Bardeen:1987vg}, the authors in~\cite{Buras:2015xba, Buras:2015yba} found:
 \begin{equation}
 Re(\epsilon'/\epsilon)_{\rm SM} =  (1.9 \pm 4.5)\times 10^{-4}\,, 
 %
 \end{equation}
 where the non-perturbative parameters at the $m_c$ scale extracted from RBC-UKQCD lattice results~\cite{Blum:2015ywa,Bai:2015nea} are taken as:
 \begin{align}
B^{(1/2)}_6(m_c)&=0.57\pm 0.19\,, \nonumber \\
 B^{(3/2)}_8(m_c) &=0.76 \pm 0.05\,.
 \end{align}
Intriguingly, although the errors in both approaches are still large,  their results are consistent with each other. If we take the results as the SM prediction,  a $2\sigma$ deviation from the experimental data of $Re(\epsilon'/\epsilon)_{\rm exp}=(16.6 \pm 2.3)\times 10^{-4}$, measured by NA48~\cite{Batley:2002gn} and KTeV~\cite{AlaviHarati:2002ye,Abouzaid:2010ny}, is indicated. 

The potential missing contributions in the SM could be from long-distance (LD) final state interactions (FSIs). However, their effects have not yet been concluded, where  the LD effects obtained  in~\cite{Buras:2016fys,Buras:2018evv}  cannot compensate for the insufficient $Re(\epsilon'/\epsilon)_{\rm SM}$, but the authors in~\cite{Gisbert:2017vvj} obtain $Re(\epsilon'/\epsilon)_{\rm SM}=(15\pm7) \times 10^{-4}$ when the short-distance (SD) and LD effects are combined. 

An alternative resolution to the small $Re(\epsilon'/\epsilon)_{\rm SM}$ is to introduce new physics effects,  such as   in~\cite{Buras:2015qea,Buras:2015yca,Buras:2015kwd,Buras:2015jaq,Tanimoto:2016yfy,Buras:2016dxz,Kitahara:2016otd,Endo:2016aws,Bobeth:2016llm,Cirigliano:2016yhc,Endo:2016tnu,Bobeth:2017xry,Crivellin:2017gks,Bobeth:2017ecx,Haba:2018byj}.  In this work, we investigate the effects of charged-Higgs-induced chromomagnetic  operators (CMOs), denoted by $\bar s\, \sigma^{\mu \nu} T^a P_{L(R)}\,d\,G^a_{\mu \nu}$, where $P_{L(R)}=(1\mp \gamma_5)/2$, $T^a$ are the $SU(3)$ generators and are normalized as $Tr(T^a T^b)=1/2 \delta^{ab}$, and $G^a_{\mu \nu}$ represents the gluon field strength tensors. It is known that the SM gluonic dipole contribution in chiral quark model could be in the region of $Re(\epsilon'/\epsilon)_{Q_{8G}} \sim (0.2,\, 0.7) \times 10^{-4}$~\cite{Buras:2015yba,Bertolini:1993rc,Bertolini:1994qk}.  Recently, a smaller $B_{CMO}$-parameter, which describes the $K\to \pi\pi$ matrix element via CMO, is consistently obtained by the lattice QCD~\cite{Constantinou:2017sgv} and a large $N_c$ dual QCD~\cite{Buras:2018evv}. Thus, it can be concluded that the SM CMO contribution cannot help to resolve the $Re(\epsilon'/\epsilon)_{\rm SM}$ problem. However, CMO induced by other new effects  may play an important role in  $Re(\epsilon'/\epsilon)$~\cite{Buras:1999da,Bertolini:2012pu}. 

We consider the  charged-Higgs contributions to CMOs  based on the following characteristics: (a)  gluon-penguin effect can be enhanced by the top-quark mass $m_t$, which arises from the top-quark Yukawa coupling  and mass insertion in the propagator; (b) a large $\tan\beta$ enhancement from Yukawa couplings could occur in the gluon-penguin diagram, and  (c) the CP violation phase can  uniquely originate from the Kobaysahi-Maskawa (KM) phase of the Cabibbo-Kobayashi-Maskawa (CKM) matrix~\cite{Cabibbo:1963yz,Kobayashi:1973fv}, where the same KM phase can be used to explain the Kaon indirect CP violation $\epsilon$ and  CP asymmetries observed in the $B$-meson system.

\begin{figure}[phtb]
\includegraphics[scale=0.75]{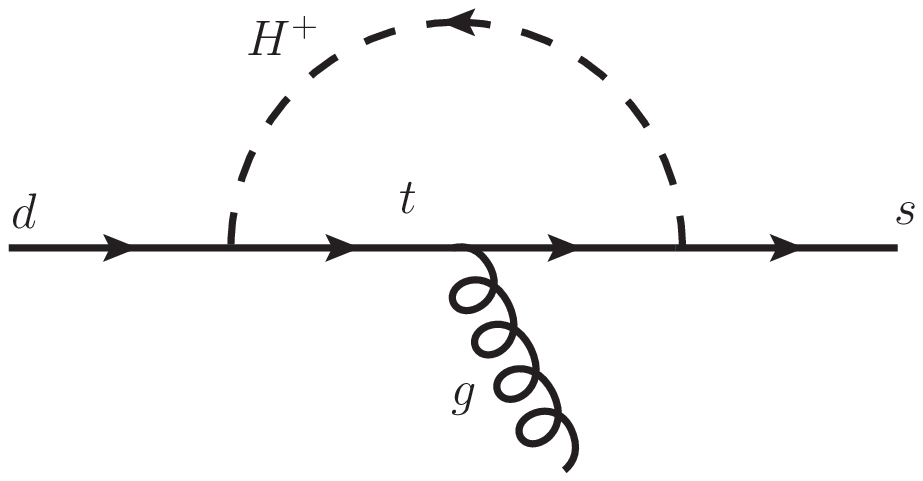}
 \caption{Sketched Feynman diagrams for $d\to s g$.  }
\label{fig:O8G}
\end{figure}

A charged-Higgs naturally exits in  a two-Higgs-doublet model (2HDM), and its Yukawa couplings strongly depend on how the Higgs doublets couple to fermions. Before discussing a specific scenario for the charged-Higgs Yukawa couplings in the 2HDM, model-independent Yukawa couplings can be generally written as:
\begin{equation}
 {\cal L}^{H^\pm}_Y =   \frac{\sqrt{2}}{v} V_{tq'} \bar t  \left( m_t C^L_{tq'} P_L  +  m_{q'} C^R_{tq'} P_R\right) q'  H^+ +H.c.,  \label{eq:tqH_general}
 \end{equation}
where $q'=d,s$; $v=\sqrt{v^2_1 + v^2_2}\approx 246$ GeV, and $v_{1(2)}$ is the vacuum expectation value (VEV) of neutral Higgs field $H_{1}(H_2)$ in doublets;  the top-quark related couplings are shown due to the $m_t$ enhancement and CP phase associated with $V_{td}$, and $C^{L(R)}_{tq'}$ denote the dimensionless  couplings to the left(right)-handed down type quarks. As a result, the effective Hamiltonian for $d\to s g$ can be expressed as:
 \begin{equation}
 {\cal H}_{d\to s g} = - \frac{G_F}{\sqrt{2}}V^*_{ts}V_{td}  \left[ y^{H^\pm}_{8G}(\mu) Q_{8G}(\mu)+  y'^{H^\pm}_{8G}(\mu) Q'_{8G}(\mu) \right] \,, \label{eq:dsH}
 \end{equation}
where the Feynman diagram is sketched in Fig.~\ref{fig:O8G}, and the dimension-6 gluonic dipole operators $Q^{(\prime)}_{8G}$ are defined as:
 \begin{align}
 Q_{8G} &= \frac{ g_s}{8 \pi^2} m_s \bar s \sigma\cdot G P_L d\,, \nonumber \\
 Q'_{8G} &= \frac{g_s }{8 \pi^2 } m_d \bar s \sigma\cdot G P_R d\,,
 \end{align}
with $g_s$ being the strong gauge coupling constant and $\sigma\cdot G = \sigma^{\mu \nu} T^a G^a_{\mu\nu}$.   The dimesionless Wilson coefficients at the $\mu_{H} \equiv \mu_{H^\pm}$ scale are obtained as:
 \begin{align}
y^{H^\pm}_{8G} (\mu_H)& =  C^{R*}_{ts} C^L_{td} J_G \left( \frac{m^2_t}{m^2_{H^\pm}}\right)\,, ~~ y'^{H^\pm}_{8G} (\mu_H) = C^{L*}_{ts} C^R_{td} J_G\left( \frac{m^2_t}{m^2_{H^\pm}}\right)\,, \\
 J_G(x) & = \frac{x ( x -3)}{(1 - x )^2} + \frac{2 x  \ln x }{(1 - x)^3}\,. \nonumber
 \end{align}
Since the electromagnetic dipole contributions are much smaller than chromomagnetic dipole, we ignore their contributions.  

To estimate the hadronic matrix element for $K^0 \to \pi \pi$ via the operators $Q^{(\prime)}_{8G}$, we take the results obtained by a dual QCD approach as~\cite{Buras:2018evv}:
 \begin{equation}
 \langle \pi \pi | C^{-}_{8G} Q_{8G}(-) | K^0\rangle \approx C^{-}_{8G}(\mu)   \frac{9}{11} \frac{m^2_\pi}{\Lambda^2_\chi} \frac{m^2_K f_\pi}{m_s(\mu) + m_d(\mu)}\approx (4.1 \times 10^{-3} \ {\rm GeV}^2) \, C^-_{8G}(\mu)\,,
 \end{equation}
where $Q_{8G}(-)\equiv g_s/(16\pi^2)\bar s \sigma^{\mu \nu} T^a \gamma_5 d G^q_{\mu \nu}$, $C^-_{8G}(\mu)$ is the effective Wilson coefficient with mass dimension $(-1)$ at the $\mu$ scale; $f_{K(\pi)}$ is the $K(\pi)$-meson decay constant, and  $\Lambda_{\chi}$ can be determined by:
 \begin{equation}
 \Lambda^2_\chi = \frac{ m^2_K - (f_K/f_\pi )m^2_\pi}{f_K/f_\pi- 1}\,,
 \end{equation}
with $f_K/f_\pi\approx 1.193$. Thus, the Kaon direct CP violation contributed by CMOs can be simply estimated as~\cite{Buras:2015yba}:
\begin{align}
 Re\left( \frac{\epsilon'}{\epsilon}\right)_{8G}  & \approx - \frac{a \omega}{\sqrt{2} |\epsilon|} \left(1-\hat{\Omega}_{\rm eff}\right) \frac{(Im A_0)_{8G}}{Re A_0} \nonumber \\
 &  \approx  - (4.1 \times 10^{-3} \ {\rm GeV}^2) \frac{a \omega (1-\hat{\Omega}_{\rm eff})}{\sqrt{2} |\epsilon| ReA_0} Im( C^-_{8G}(\mu))  \,,  \label{eq:epsilon_p1} \\
 C^-_{8G}(\mu) & = - \frac{G_F}{ \sqrt{2}} V^*_{ts} V_{td} \left(m_d y'^{H^\pm}_{8G} (\mu) - m_s y^{H^\pm}_{8G} (\mu)\right)\,,
 \label{eq:epsilon_p2}
  \end{align}
where $a=1.017$~\cite{Cirigliano:2003gt} and $\hat \Omega_{\rm eff}=(14.8 \pm 8.0)\times 10^{-2}$~\cite{Buras:2015yba} include the isospin breaking corrections and the correction of $\Delta I=5/2$; $\omega=Re A_2/Re A_0\approx 1/22.46$ denotes the $\Delta I=1/2$ rule, and $A_{0(2)}$ denotes the $K^0\to \pi^+ \pi^-$ decay amplitude in the isospin $I=0(2)$ final state. With $|\epsilon|\approx 2.228\times 10^{-3}$~\cite{PDG} and $ReA_0 \approx 27.12 \times 10^{-8}$ GeV, Eq.~(\ref{eq:epsilon_p1}) can be expressed as:
 \begin{equation}
 Re\left( \frac{\epsilon'}{\epsilon}\right)_{8G}   \approx - (1.85 \times 10^{5} \ {\rm GeV}) \times Im( C^-_{8G}(\mu))  \,.  \label{eq:epsilon_p3}
 \end{equation} 

After formulating the necessary part for the calculation of $Re(\epsilon'/\epsilon)_{8G}$, in the following, we discuss $C^{L(R)}_{tq'}$ in the 2HDM. In the literature, to avoid flavor-changing neutral currents (FCNCs) at the tree level, usually  global symmetry is typically  imposed  in the 2HDM. According to the different symmetry transformations, the 2HDM can be classified as type-I, type-II, lepton-specific, and flipped models, for which a detailed discussion can be found in~\cite{Branco:2011iw}. The dimensionless parameters $C^L_{tq'}$ and $C^R_{tq'}$ in these models  can be found as in~\cite{Branco:2011iw}:
\begin{align}
C^L_{ts}&=C^L_{td} =\cot\beta\,, ~ C^R_{td}=C^R_{ts}=-\cot\beta\,, ~~~\text{type-I \& lepton-specific (Type A)} \,, \nonumber \\
C^L_{ts} & = C^L_{td} = \cot\beta\,, ~ C^R_{ts}=C^R_{td}= \tan\beta\,, ~~~~~\text{type-II \& flipped (Type B)}\,,
\end{align}
with $\tan\beta=v_2/v_1$. 
As a result, the direct CP violation can be estimated as:
 \begin{equation}
 Re\left(\frac{\epsilon'}{\epsilon} \right)_{8G} \approx  1.07 \times 10^{-5}   \left\{\begin{array}{cc}
   -\cot^2\beta \frac{J_{G}( m^2_t/m^2_{H^\pm})}{-0.45}  & \text{(Type A)}\,, \\ 
    \frac{J_{G}( m^2_t/m^2_{H^\pm})}{-0.45} & \text{(Type B)}\,,\\ 
  \end{array} \right.
 \end{equation}
where we have used $m_s(m_c) \approx 0.109$ GeV, $m_d(m_c)\approx 5.44$ MeV~\cite{Buras:2015yba}, and $m_{H^\pm}=300$ GeV to estimate the numerical values. From the analysis, it can be clearly seen that the contributions of CMOs in the type-II and flipped models are independent of $\tan\beta$, and the magnitude is two orders of magnitude smaller than the data. The situation in type-I and lepton-specific models is worse, where even the sign is opposite.  It is obvious that it is necessary to look for another scenario to enhance $Re(\epsilon'/\epsilon)$ in the 2HDM. 

The most feasible scenario is the use of the  generic 2HDM without imposing extra global symmetry, i.e. the type-III 2HDM, where the model can be successfully used to resolve  anomalies that are indicated by the experiments, such as $h\to \mu \tau$, muon $g-2$, $R(D)$, and $R(K^{(*)})$~\cite{Benbrik:2015evd,Chen:2017eby,Akeroyd:2017mhr,Arhrib:2017yby,Chen:2018hqy}. Although the type-III 2HDM has  tree-level FCNCs, the flavor-changing effects, which involve light quarks, can be naturally suppressed when the Cheng-Sher ansatz~\cite{Cheng:1987rs} is applied. In order to understand the new  characteristics  of the charged-Higgs interactions, we write the  $H^\pm$ Yukawa couplings to the quarks in the type-III model as~\cite{Benbrik:2015evd,Chen:2017eby,Chen:2018hqy}:
\begin{align}
-{\cal L}^{H^\pm}_Y &=  \sqrt{2} \bar u_R   \left[ - \frac{ 1}{v \tan\beta } {\bf m_u} + \frac{{\bf X}^{u\dagger}}{ \sin\beta}  \right] V d_L  H^+  \nonumber \\ 
&+ \sqrt{2} \bar u_L V \left[ - \frac{ \tan\beta}{v } {\bf m_d} + \frac{{\bf X}^d}{  \cos\beta}  \right] d_R H^+   + H.c.\,, \label{eq:CHY}
\end{align}
where the flavor indices are suppressed; $V\equiv V^u_L V^{d^\dagger}_L$ stands for the CKM matrix, and ${\bf X^{u,d}}$ are defined as:
\begin{align}
 {\bf X^u}= V^u_L \frac{Y^u_1}{\sqrt{2}} V^{u\dagger}_R \,, \     {\bf X^d}= V^d_L \frac{Y^d_2}{\sqrt{2}} V^{d\dagger}_R \,. \label{eq:Xfs}
\end{align}
Here, $V^{u,d}_{L,R}$ are the unitary matrices for diagonalizing the quark mass matrices, and $Y^{u(d)}_{1(2)}$ is one of two Yukawa matrices, which consist of the  up(down)-type quark mass matrix. When $Y^{u(d)}_{1(2)}$ vanishes, the Yukawa couplings in the type-II model will be recovered. Thus, the new effects are from ${\bf X^u}$ and ${\bf X^d}$, and they indeed dictate the  tree-level FCNC effects. Using the Cheng-Sher ansatz, we can parametrize the ${\bf X^{u,d}}$ as:
 \begin{align}
 %
 {\bf X^f}_{ij} & = \frac{\sqrt{m_{f_i} m_{f_j} }}{v} \chi^f_{ij}\,. \label{eq:CSA}
 \end{align}
In terms of above parametrizations, the new parameters $\chi^u_{ij}$ and $\chi^d_{ij}$ in general can be of $O(1)$, and their magnitudes can be determined or constrained by the experimental data. 

It is of interest to understand why the  ${\bf X^{u,d}}$ effects can play an important role in flavor physics. From Eqs.~(\ref{eq:CHY}) and (\ref{eq:CSA}), the vertices of $t_R q'_L H^+$ can be expressed as:
 \begin{equation}
\frac{\sqrt{2} }{\sin\beta}\left[{\bf X}^{u\dagger} V\right]_{tq'} = \frac{\sqrt{2} m_t}{v \sin\beta} \sum_{j=u,c,t}\sqrt{\frac{m_{u_j }}{m_t}} \chi^{u*}_{jt} V_{jd'}\,.
 \end{equation}
Due to $\sqrt{m_u/m_t} V_{uq'} < \sqrt{m_c/m_t} V_{cq'}$, we can simplify above equation as:
 \begin{equation}
\frac{\sqrt{2} }{\sin\beta}\left[{\bf X}^{u\dagger} V\right]_{tq'} \approx \frac{\sqrt{2} m_t}{v \sin\beta}  \left( \chi^{u*}_{tt} V_{tq'} + \sqrt{\frac{m_c}{m_t}} \chi^{u*}_{ct} V_{cq'} \right)\,.
 \end{equation}
 Using $|V_{ts(td)}|\approx 0.041(0.0088)$ and $|V_{cs(cd)}|\approx 1 (0.225)$, $\sqrt{m_c/m_t} V_{cq'}/V_{tq'}$ can be estimated  to be $\sim 2.27$ for $q'=d$ and $\sim 2.14$ for $q'=s$; that is, the second term can have an important effect. In addition, for a  large $\tan\beta$ scheme, the vertex of $t_R q'_R H^\pm$ is insensitive to $\tan\beta$.  Similarly, the vertex of $t_L q' H^+$ can be simplified as:
  \begin{equation}
  \frac{\sqrt{2}}{\cos\beta} \left[ V {\bf X}^d \right]_{tq'} \approx \frac{\sqrt{2} m_b }{v \cos\beta} V_{tb} \sqrt{\frac{m_{q'}}{m_b}}\chi^d_{bq'}
  \end{equation}
 where we only retain the term with maximal CKM matrix element $V_{tb}\approx 1$.  Intriguingly, the vertex of $t_L q'_R H^+$ does not have the $V_{tq'}$ suppression factor and its dependence on $m_{q'}$ is smeared by the square-root of $m_{q'}$. Unlike the case of $t_R q'_L H^+$, the $t_L q'_R H^+$ coupling is sensitive to $\tan\beta$. 
 
From Eqs.~(\ref{eq:CHY}) and (\ref{eq:CSA}), we can write the relevant charged-Higgs couplings  as:
 \begin{align}
 {\cal L}^{H^\pm}_Y & \supset  \frac{\sqrt{2}}{v} V_{tq'} \bar t  \left( m_t \zeta^u_{tq'} P_L  -  m_b \zeta^d_{tq'} P_R\right) q'  H^+ +H.c.,  \label{eq:tqH}
 \end{align}
where  the parameters $\zeta^{f}_{tq'}$ $(q'=d,s)$ can be of $O(1)$  and  are defined as:
\begin{align}
\zeta^u_{tq'} & = \frac{1}{\tan\beta} - \frac{\chi^L_{tq'} }{\sin\beta}\,, \ \chi^L_{tq'}  =\chi^{u*}_{tt} + \sqrt{\frac{m_c}{m_t}} \frac{V_{cq'}}{V_{tq'}} \chi^{u*}_{ct}\,, \nonumber \\
\zeta^{d}_{tq'} & = \tan\beta  \sqrt{\frac{m_{q'}}{m_b}} \frac{V_{tb}}{V_{tq'}}  \frac{\chi^d_{bq'}}{\sin\beta} \,.
  \label{eq:zetas}
\end{align}
According to the expressions in Eqs.~(\ref{eq:dsH}) and~(\ref{eq:epsilon_p2}), the Wilson coefficients for CMOs at the $\mu_H$ scale in the type-III model can be written as:
 \begin{align}
 m_s y^{H^\pm}_{8G}(\mu_H) & = -m_b\zeta^{d*}_{ts} \zeta^u_{td} J_{G}\left(\frac{m^2_t}{m^2_{H^\pm}}\right)\,,
  \nonumber \\
m_d  y'^{H^\pm}_{8G}(\mu_H) & =-m_b \zeta^{u*}_{ts} \zeta^d_{td} J_{G}\left(\frac{m^2_t}{m^2_{H^\pm}}\right)\,.
 \end{align}
If the source of CP violation in the type-III 2HDM still originates from the KM phase, we find that the imaginary part of the effective Wilson coefficient $C^-_{8G}(\mu_H)$  is only related to $m_s y^{H^\pm}_{8G}$ and has a simple form as follows:
\begin{equation}
Im(C^-_{8G}(\mu_H))= \sqrt{2} m_b G_F Im(V^*_{ts} V_{td}) \zeta^d_{ts} \left( \frac{1}{\tan\beta} - \frac{\chi^u_{tt}}{\sin\beta}\right)\,. \label{eq:muHC8G}
\end{equation}
 In addition to the $m_{H^\pm}$ and $\tan\beta$ parameters, the  new parameters for $Re(\epsilon'/\epsilon)_{8G}$ involved in the type-III 2HDM are only $\chi^d_{bs}$ and $\chi^u_{tt}$, where $|\chi^d_{bs}|$ could be of $O(10^{-2})$ due to the constraints from flavor physics, and $|\chi^u_{tt}|$ could be of $O(1)$~\cite{Chen:2018hqy,Chen:2018ytc}. In addition, because of the $\chi^u_{tt}$ term, the $\tan\beta$ enhancement factor from $\zeta^d_{ts}$ is retained, so the $Re(\epsilon'/\epsilon)_{8G}$ can be significantly enhanced in a large $\tan\beta$ scheme. Note that although Eq.~(\ref{eq:muHC8G}) is shown  at the $\mu_H$ scale, in our numerical analysis, we take the value at the $\mu=m_c$ scale using the renormalization group (RG) running.

Before discussing the numerical analysis, we discuss the  parameters that are involved as well as the theoretical and experimental inputs.  In addition to $Re(\epsilon'/\epsilon)_{8G}$,  the charged-Higgs has  significant contributions on the low energy flavor physics, such as the $\Delta B_{q'}=2$, $\Delta K=2$, $B\to X_s \gamma$, $B_s \to \mu^+ \mu^-$ processes, and the indirect CP violation $\epsilon$~\cite{Chen:2018hqy,Chen:2018ytc}.  Although the $Re(\epsilon'/\epsilon)_{8G}$ related parameters are only $\tan\beta$, $m_{H^\pm}$, $\chi^u_{t}$, and $\chi^d_{bs}$, since these parameters and the other parameters  are correlated  in the mentioned flavor physics, we have to consider all parameters together when the strictly experimental constraints are taken into account. Therefore,   the involved parameters and their taken ranges are shown as:
\begin{align}
&  -1 \leq  \chi^d_{bb}   \leq 1\,,  ~ -0.08 \leq \chi^d_{bd} \leq 0.08\,, ~ 0 \leq \chi^u_{tt} \leq 1\,, ~ -1 \leq \chi^u_{ct} \leq 0\nonumber \\
 &   0\leq \chi^d_{bs} \leq 0.08\,, ~\tan\beta\in (20,\, 50)\,, ~ m_{H^\pm} \in (200,\, 500)\, \text{GeV}\,, \label{eq:para}
 \end{align}
 where we choose $\chi^u_{tt}$ and $\chi^u_{ct}$ to be opposite in sign because the resulted $Re(\epsilon'/\epsilon)$ from the gluon and electroweak penguins can reach $O(10^{-4})$~\cite{Chen:2018ytc}. 
 
 In addition to the charged-Higgs effects, the neutral scalars $H$ and $A$ in the type-III model can also contribute to the relevant flavor physics via the tree-level FCNC effects. It was found that  with the exception of the tree-level contributions to $\Delta B(K)=2$, the  loop-induced $\Delta B=1$ processes through the mediation of $H(A)$  boson are  suppressed by $m^2_b/m^2_{H^\pm}$ and can be neglected~\cite{Chen:2018ytc}. Since the same suppression also appears in the CMOs,  we thus neglect the $H(A)$ contributions to $Re(\epsilon'/\epsilon)_{8G}$.  In addition, although the tree-level flavor-changing couplings $sdH(A)$ can also contribute to $K\to \pi \pi$, due to the suppression factor $m_{d(u)} \tan\beta/v$ and the constraint from $\Delta S=2$,  their effects are small and negligible. Hence, in order to include the tree-level $H(A)$-mediated contributions to $\Delta B=2$, which can give strict constraints on $\chi^{d}_{bs, bd}$, we simply fix $m_H=m_A=600$ GeV. 

The experimental inputs   used to bound the free parameters are taken as~\cite{PDG,Aaij:2017vad}:
   \begin{align}
  \Delta M^{\rm exp}_K & \approx 3.48 \times 10^{-15} \text{ GeV} \,,  \ 
     \Delta M^{\rm exp}_{B_d}  =(3.332 \pm 0.0125)\times 10^{-13}\, \text{GeV}\,, \nonumber \\
    \Delta {M^{\rm exp}_{B_s}} & =(1.168 \pm 0.014) \times 10^{-11}\, \text{GeV}\,, \    BR(B\to X_s \gamma)^{\rm exp}   = (3.49 \pm 0.19)\times 10^{-4}\,, \nonumber \\
\epsilon^{\rm exp}  & \approx 2.228\times 10^{-3}\,, \  BR(B_s \to \mu^+ \mu^-)^{\rm exp}   = (3.0\pm 0.6 ^{+3.0}_{-2.0})\times 10^{-9}\,. 
  \end{align}
Since $\epsilon$ in the SM fits well with the experimental data~\cite{Buchalla:1995vs}, for new physics effects, we thus use~\cite{Buras:2015jaq}:
 \begin{equation}
| \epsilon^{\rm NP}| \leq  0.4 \times 10^{-3}\,.  \label{eq:eNP_K}
 \end{equation} 
For the $\Delta K=2$ process, we take a combination of  the short-distance (SD) and long-distance (LD) effects in the SM as $\Delta M^{\rm SM}_K (SD + LD)    = (0.80 \pm 0.10) \Delta M^{\rm exp}_K$~\cite{Buras:2014maa}; therefore,   the allowed new physics contribution to $\Delta M_K$ should satisfy: 
  \begin{equation}
   |\Delta M^{\rm NP}_K|\leq  0.2 \Delta M^{\rm exp}_K \,. \label{eq:DMNP_K}
  \end{equation}
The values of the CKM matrix elements are taken as:
 \begin{align}
 V_{ud} & \approx V_{cs} \approx 0.973\,, \ V_{us}\approx - V_{cd}\approx 0.225\,, \nonumber \\
 V_{td} & \approx 0.0088 e^{-i \phi_2}\,, \ \phi_2 \approx 23^{\circ}\,, \ V_{ts} \approx -0.041 \,, \ V_{tb}\approx 1\,, 
 \end{align}
 where $Re(V^*_{ts} V_{td}) \approx  -3.3\times 10^{-4}$ and $Im(V^*_{ts} V_{td}) \approx 1.4 \times 10^{-4}$ are taken to be the same as those used in~\cite{Buras:2015yba}. The particle masses used to estimate the numerical values are given as:
  \begin{align}
  & m_K\approx 0.489 \text{ GeV}\,,~m_{B_d}\approx 5.28 \text{ GeV}\,,~m_{B_s}\approx 5.37 \text{ GeV}\,,~m_W \approx 80.385 \text{ GeV} \,, \nonumber \\
  & m_t \approx 165 \text{ GeV}\,, ~m_c\approx 1.3 \text{ GeV}\,, ~m_s(m_c) \approx  0.109 \text{ GeV}\,, ~m_d (m_c) \approx 5.44 \text{ MeV}\,.
  \end{align}

To consider the constraints from $\Delta M_{B_q'}$, $B \to X_s \gamma$, $\Delta M_K$, and $\epsilon$, we employ  the results  obtained in~\cite{Chen:2018hqy,Chen:2018ytc}. In the generic 2HDM, the $B_s \to \mu^+ \mu^-$ process can arise from the charged-Higgs induced $Z(A)$-penguin and  box diagrams and the tree-level $bsA$ FCNC associated with the $A\to \mu^+ \mu^-$. Since the $H^\pm(A)$ Yukawa couplings to the leptons are dictated by $1- \chi^\ell_\ell/\sin\beta$, the $B_s \to \mu^+ \mu^-$ process, which is generated by the pseudoscalar $A$ and  the charged-Higgs box diagrams, can be suppressed when $\chi^\ell_\mu \approx 1$ is taken. As a result, the dominant contribution to $B_s \to \mu^+ \mu^-$ is from the $H^\pm$-induced $Z$-penguin diagrams, and the branching ratio  for the $B_s \to \mu^+ \mu^-$ decay can be obtained as:
\begin{equation}
BR(B_s \to \mu^+ \mu^-) = BR(B_s \to \mu^+ \mu^-)^{\rm SM} \left| 1 + \frac{C^{H^\pm}_{10} }{C^{\rm SM}_{10} }\right|^2\,,
\end{equation}
where $C^{\rm SM}_{10} \approx - 4.21$~\cite{Buchalla:1995vs} denotes the SM contribution, and the charged-Higgs contribution with $x_q=m^2_q/m^2_W$ and $y_t=m^2_t/m^2_{H^\pm}$ is given as:
 \begin{align}
 C^{\rm H^\pm}_{10} & = -y_t \left( g^u_L \zeta^{u*}_{ts} \zeta^u_{tb} x_t - g^u_R \zeta^{d*}_{ts} \zeta^d_{tb} x_b \right) J_1(y_t)\,, \nonumber \\
 g^u_L & = \frac{1}{2} - \frac{2}{3}\sin^2\theta_W\,, \ g^u_R = - \frac{2}{3} \sin^2\theta_W\,, \nonumber \\
 \zeta^u_{tb} &= \frac{1}{\tan\beta} - \frac{\chi^{u*}_{tt}}{\sin\beta}\,, \ \zeta^d_{tb}  = \tan\beta(1- \chi^d_{bb})\,, \nonumber \\
 J_1(y_t) &= -\frac{1}{4} \left(\frac{1}{1-y_t} + \frac{\ln y_t}{(1-y_t)^2}\right)\,.
 \end{align}
 Due to the suppression factor of $x_b=m^2_b/m^2_W$ in the second term of $C^{H^\pm}_{10}$, it can be seen that $B_s \to \mu^+ \mu^-$ cannot give a strict limit on $\chi^d_{bs}$. 
 
\begin{figure}[phtb]
\includegraphics[scale=0.6]{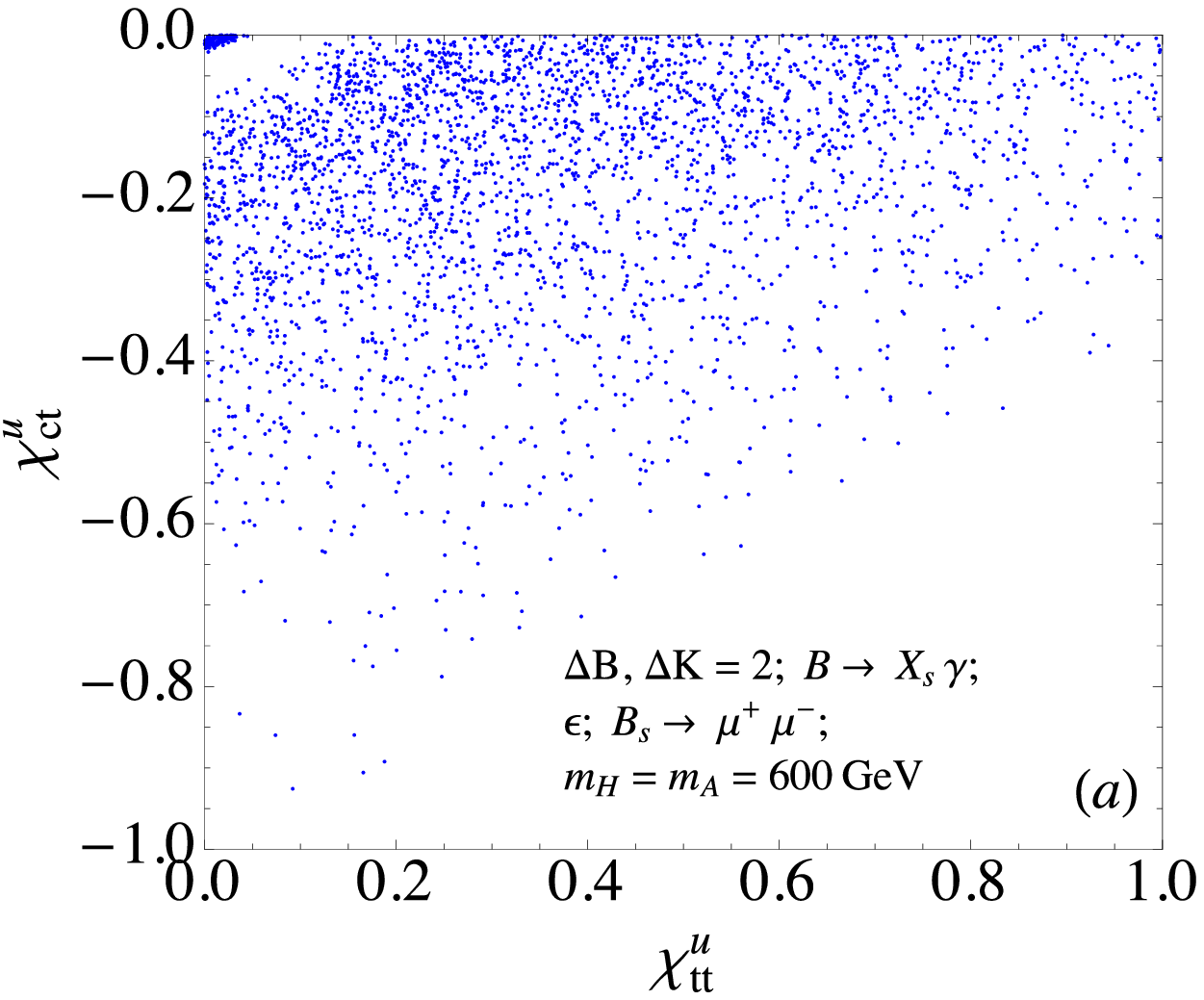}
\includegraphics[scale=0.6]{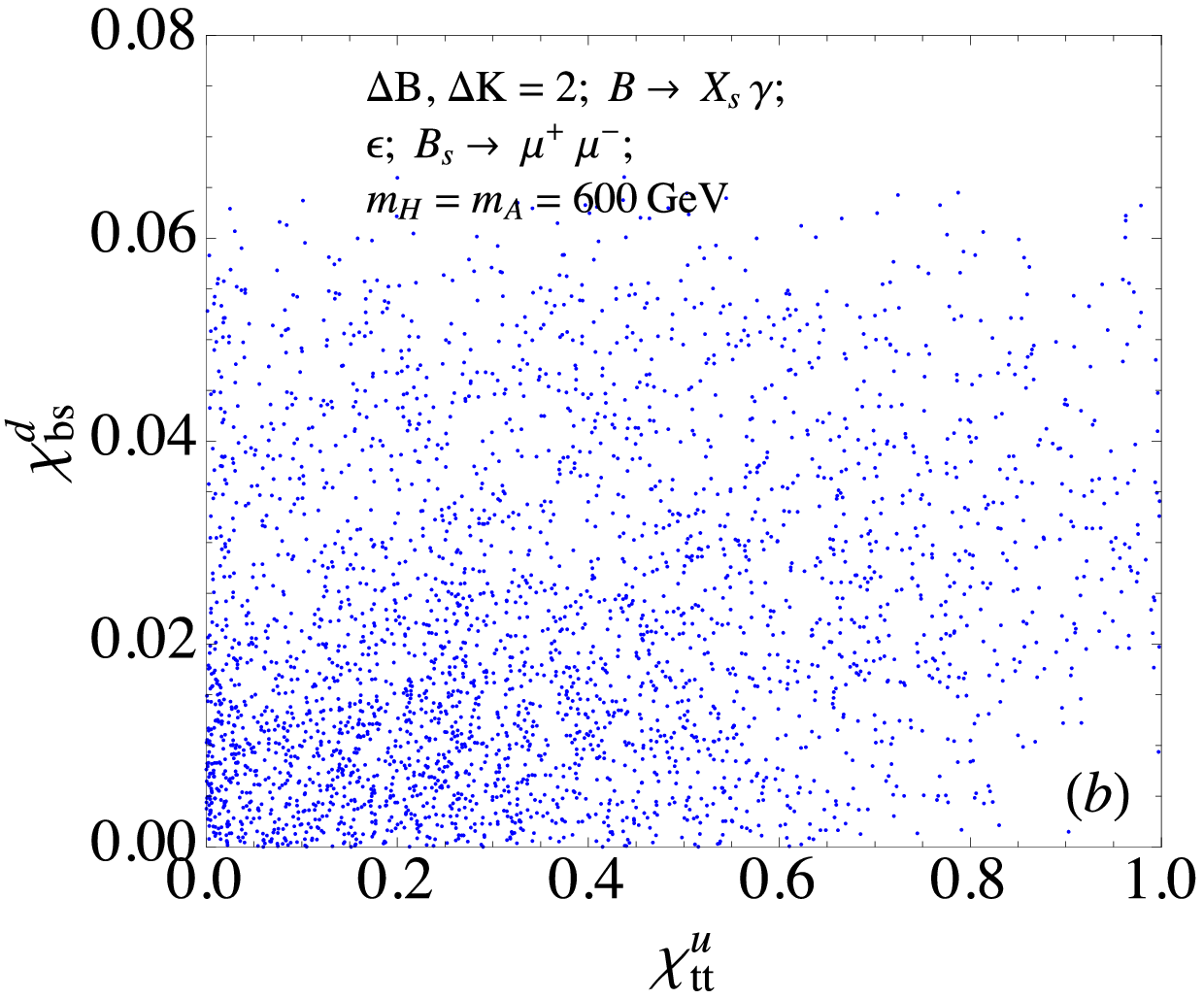}
 \caption{Allowed parameter spaces for (a) $\chi^u_{tt}$-$\chi^u_{ct}$ and (b) $\chi^u_{tt}$-$\chi^d_{bs}$, where the number of sampling points used for the scan is $5\times 10^6$. }
\label{fig:limit}
\end{figure}

\begin{figure}[phtb]
\includegraphics[scale=0.7]{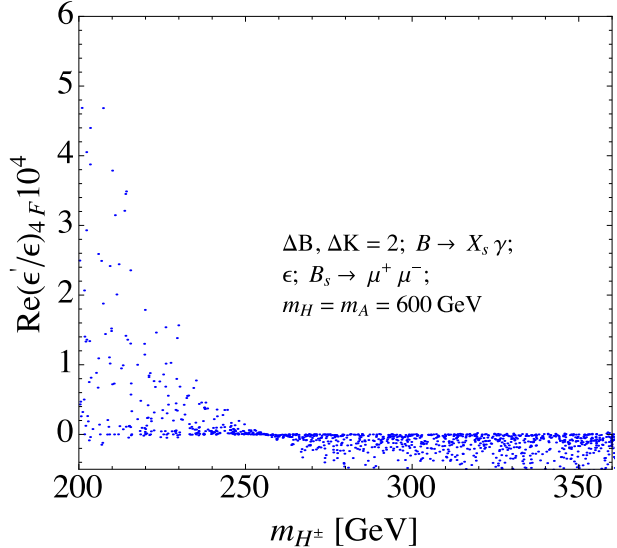}
 \caption{$Re(\epsilon'/\epsilon)_{4F}$  from the gluon and electroweak penguin four-fermion operators as a function of $m_{H^\pm}$, where the constraints from the experimental data are included.  }
\label{fig:4F}
\end{figure}

 Since the uncertainty  of  $BR(B_s \to \mu^+ \mu^-)^{\rm SM}=(3.65 \pm 0.23)\times 10^{-9}$~\cite{Bobeth:2013uxa} is smaller than  the errors of the LHCb result, to consider the  $B_s \to \mu^+ \mu^-$ constraint, we require that $C^{H^\pm}_{10}/C^{\rm SM}_{10}$ is  less than $2\sigma$ of the experimental value, i.e., $|C^{H^\pm}_{10}/C^{\rm SM}_{10}| \lesssim 0.2$. Using the experimental and theoretical inputs mentioned earlier, the bounds on $\chi^u_{tt}$-$\chi^u_{ct}$ and $\chi^u_{tt}$-$\chi^d_{bs}$ are shown in Fig.~\ref{fig:limit}(a) and (b), where we use $5\times 10^{6}$ sampling points to scan the parameters. From the results, we see  that  $|\chi^u_{tt}|\lesssim  1.0$ and  $|\chi^d_{bs}| \lesssim 0.06$ are allowed  in the chosen region of Eq.~(\ref{eq:para}) .  
Using the constrained parameters, we can estimate the Kaon direct CP violation derived from the gluon and electroweak penguin operators. According to the result obtained in~\cite{Chen:2018ytc}, the $Re(\epsilon'/\epsilon)_{4F}$ via the penguin four-fermion operators as a function of $m_{H^\pm}$ is given in Fig.~\ref{fig:4F},  where the electroweak penguin operator $Q_8$ dominates. From the result, it can be seen that the penguin four-fermion operator contribution becomes $Re(\epsilon'/\epsilon) _{4F}< 10^{-4}$ when $m_{H^\pm} \gtrsim 230$ GeV.

\begin{figure}[phtb]
\includegraphics[scale=0.5]{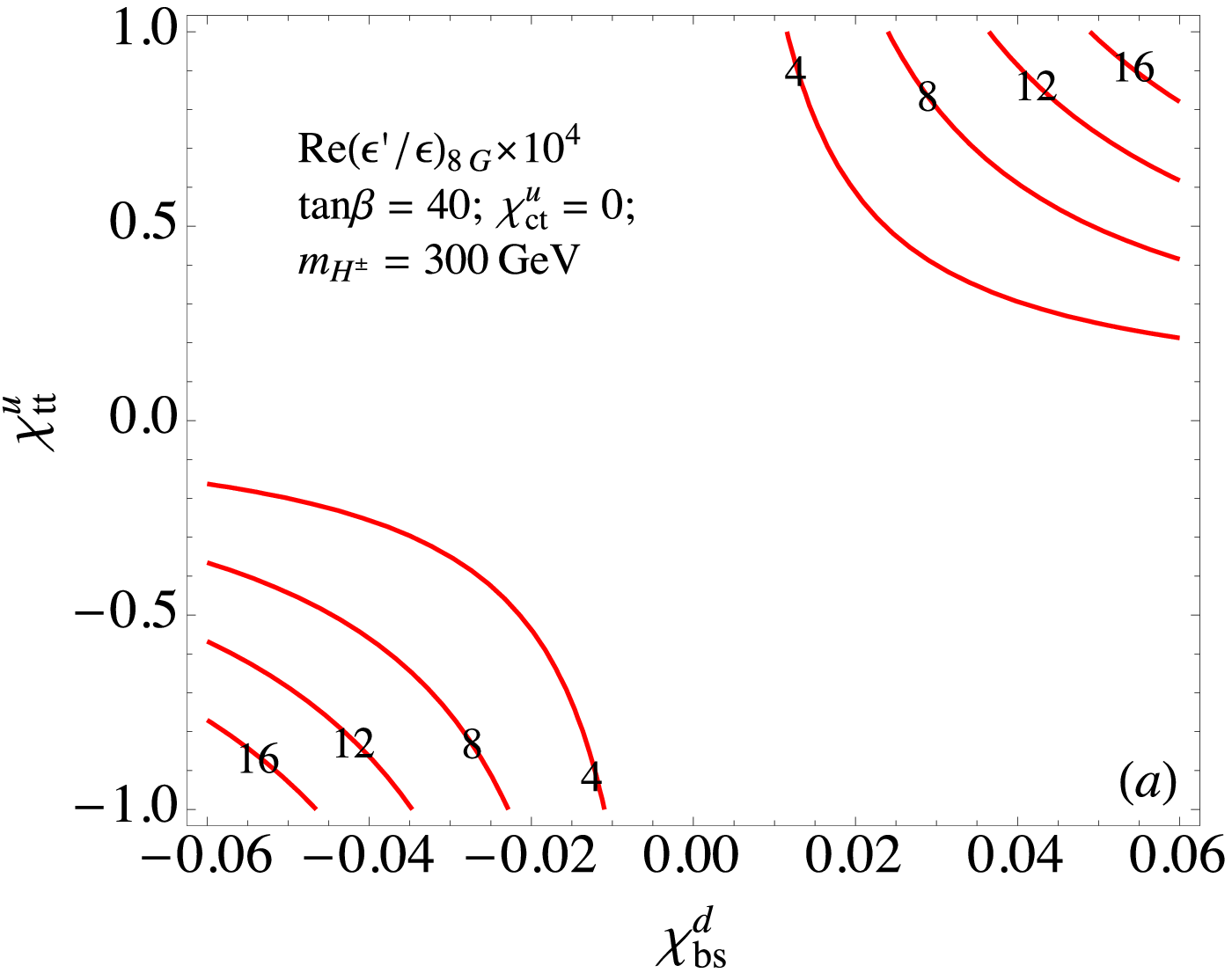}
\includegraphics[scale=0.5]{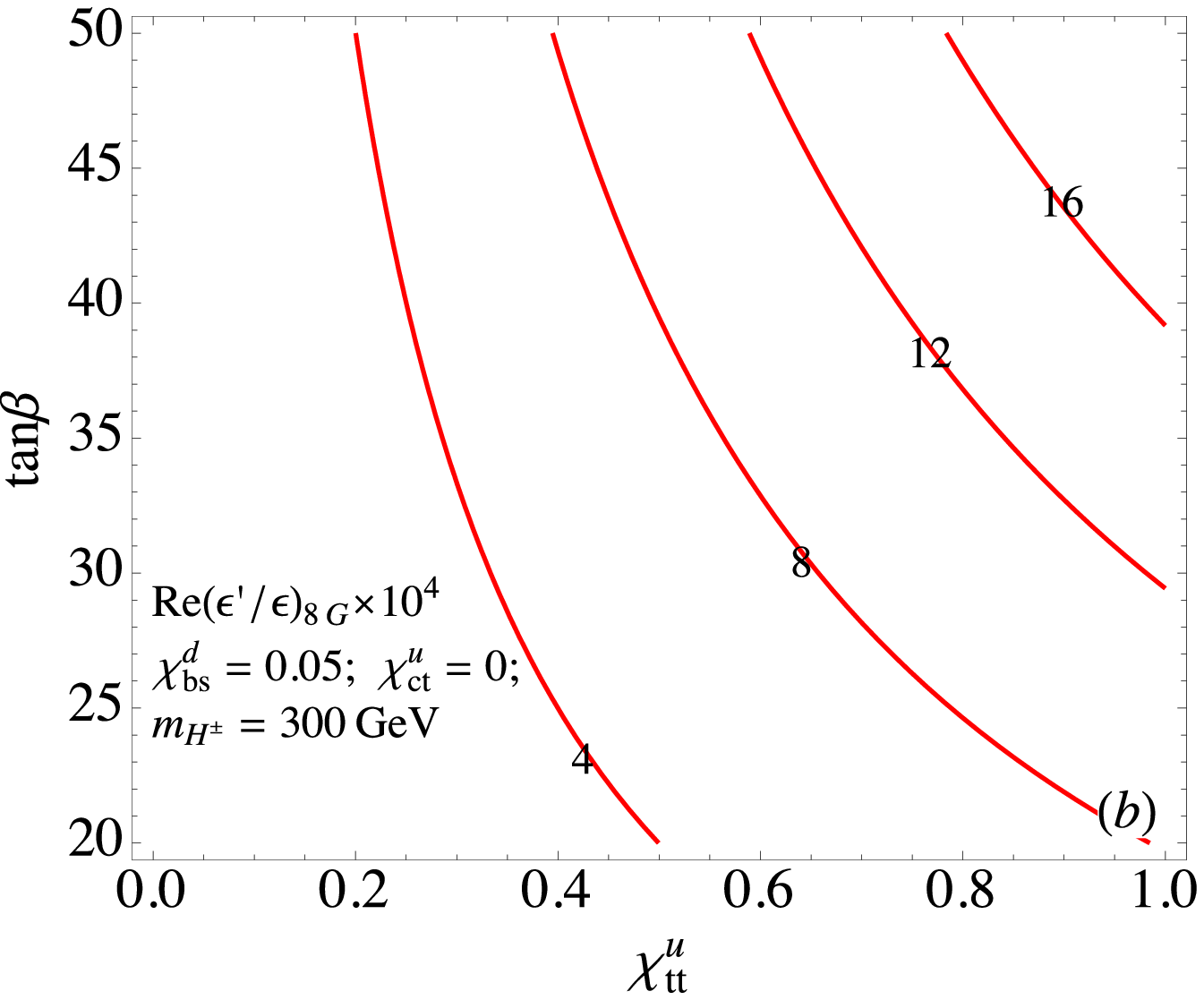}
\includegraphics[scale=0.5]{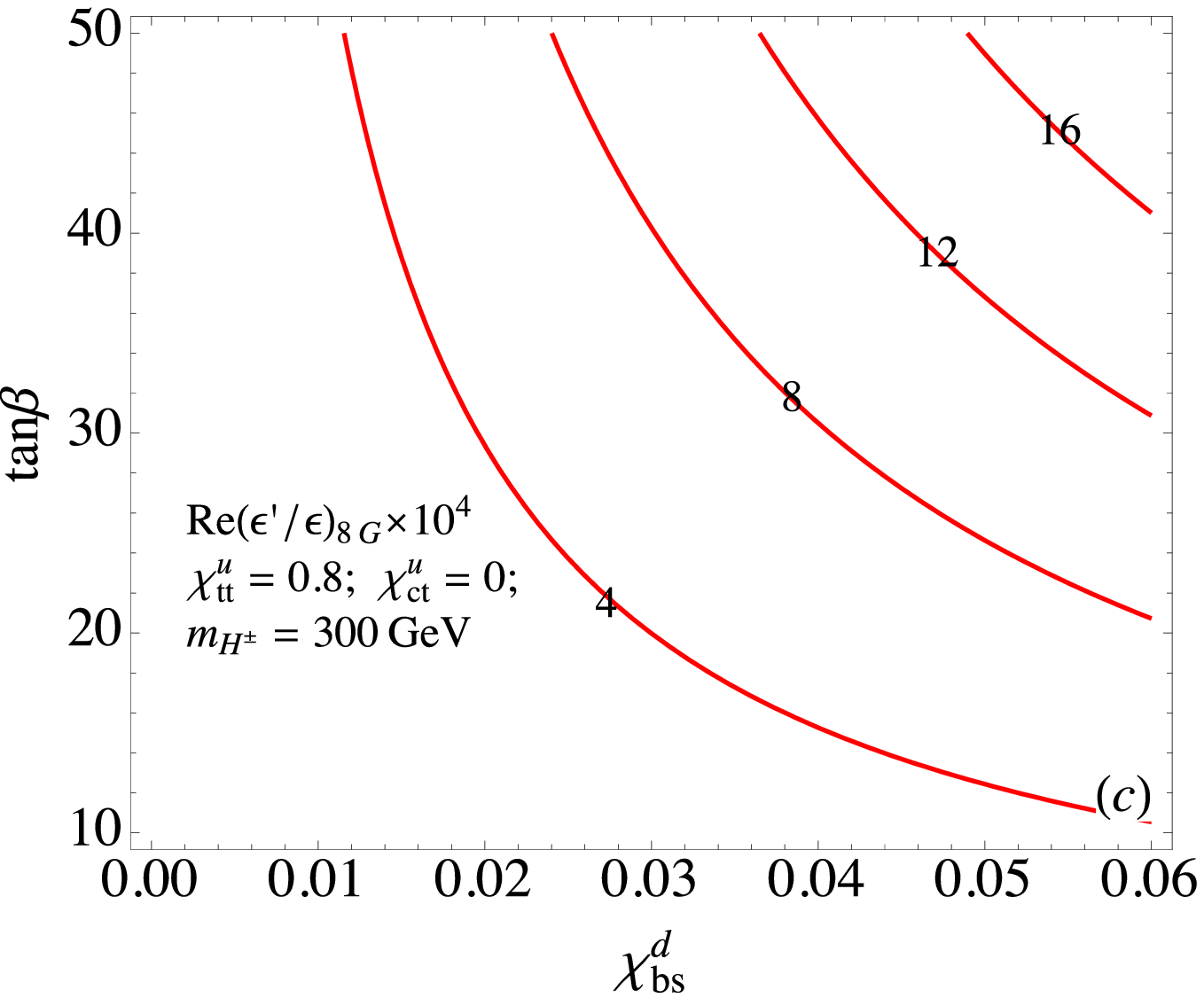}
\includegraphics[scale=0.5]{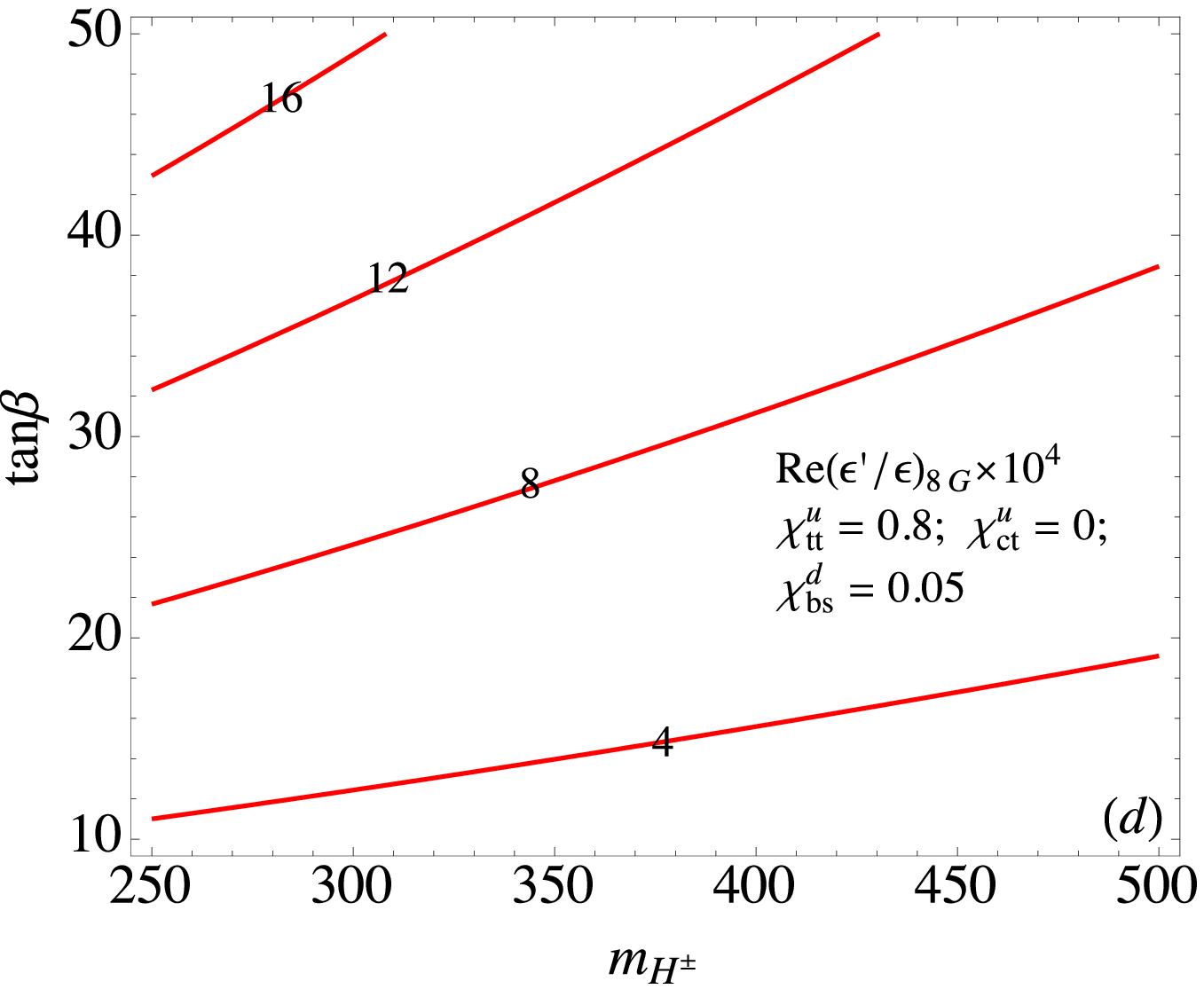}
 \caption{$Re(\epsilon'/\epsilon)_{8G}$ (in units of $10^{-4}$)  as a function of (a)  $\chi^{d}_{bs}$ and $\chi^u_{tt}$, (b) $\chi^u_{tt}$ and $\tan\beta$, (c) $\chi^d_{bs}$ and $\tan\beta$, and (d) $m_{H^\pm}$ and $\tan\beta$, where the values of the fixed parameters are shown in the plots. }
\label{fig:O8G_ad}
\end{figure}

In the following, we discuss the CMO contribution to $Re(\epsilon'/\epsilon)$. Since only $y^{H^\pm}_{8G}(\mu_H)$ contributes to the Kaon direct CP violation in this model, the associated Wilson coefficient at the $\mu=m_c$ scale can be obtained as:
 \begin{equation}
 y^{H^\pm}_{8G} (m_c) \approx  -0.137 C_2 (m_W) + 0.487 y^{H^\pm}_{8G}(\mu_H)\,,
 \end{equation}
 where $C_2(m_W)\approx 1$ is the Wilson coefficient of the $Q_2=(\bar s c)_{V-A} (\bar c d)_{V-A}$ operator; for the new physics effects, we only use the leading-order QCD anomalous-dimension matrix for the operators $Q_{1-6}$, $O_{7\gamma}$, and $Q_{8G}$~\cite{Buchalla:1995vs}, and $\mu_H=300$ GeV is fixed. Using Eqs.~(\ref{eq:epsilon_p3}) and (\ref{eq:muHC8G}), we show $Re(\epsilon'/\epsilon)_{8G}$ in units of $10^{-4}$ as a function of $\chi^d_{bs}$-$\chi^u_{tt}$ in Fig.~\ref{fig:O8G_ad}(a), where $\tan\beta=40$ and $m_{H^\pm}=300$ GeV are fixed. Since the results are not sensitive to the $\chi^u_{ct}$ parameter,  we take $\chi^u_{ct}=0$ in our numerical estimates. To obtain a positive $Re(\epsilon'/\epsilon)_{8G}$, $\chi^d_{bs}$ and $\chi^u_{tt}$ have to be the same sign.  We show the correlations of $Re(\epsilon'/\epsilon)_{8G}$ to $\chi^u_{tt}$-$\tan\beta$ and $\chi^d_{bs}$-$\tan\beta$ in plots~Fig.~\ref{fig:O8G_ad}(b) and (c), respectively, where we only show the positive $\chi^u_{tt}$ and $\chi^d_{bs}$ because the results for the case of $\chi^{u(d)}_{tt(bs)} < 0$ are the same. The correlation of $Re(\epsilon'/\epsilon)_{8G}$ to $m_{H^\pm}$ and $\tan\beta$  is shown in Fig.~\ref{fig:O8G_ad}(d). From these results, it can be clearly seen that with the constrained parameter values from the low energy flavor physics,  even in the case of  $m_{H^\pm} > 250$ GeV, the Kaon direct CP violation from the charged-Higgs-induced CMO in the type-III 2HDM can easily reach  the level of  $O(10^{-3})$.

In summary, we studied the Kaon direction CP violation $Re(\epsilon'/\epsilon)$, which arises from the charged-Higgs-induced chromomagnetic operator. If we assumed that  the Kobayashi-Maskawa phase is the unique origin of CP violation, it was found that in addition to the $\tan\beta$ parameter, the new parameters $\chi^u_{tt}$ of $O(1)$ and $\chi^d_{bs}$ of $O(10^{-2})$ in the type-III 2HDM play an important role in the contribution of the gluon dipole, where  the former  is a flavor-conserving coupling, and the latter is directly associated with the tree-level FCNCs. Although the contributions of the charged-Higgs-induced gluon and electroweak penguin operators  to $Re(\epsilon'/\epsilon)$ can be $\sim 5 \times 10^{-4}$ when $m_{H^\pm} \sim 200$ GeV, it was found that the charged-Higgs-induced chromomagnetic operator  can explain the observed $Re(\epsilon'/\epsilon)$ with wider parameter spaces, even in the case of  $m_{H^\pm} > 250$ GeV.

\section*{Acknowledgments}

This work was partially supported by the Ministry of Science and Technology of Taiwan,  
under grants MOST-106-2112-M-006-010-MY2 (CHC).

\end{document}